\documentclass{WileyMSP-template}
\usepackage{amsmath}
\usepackage{siunitx}

\begin{document}

\justifying 
\setlength{\parindent}{0pt}
\setlength{\parskip}{0.6em}

\pagestyle{fancy}
\rhead{\includegraphics[width=2.5cm]{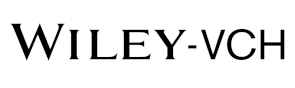}}

\title{Dimple-Encoded Reprogrammable Origami}

\maketitle


\author{Qun Zhang$^1$,}
\author{Weicheng Huang$^{2,*}$,}
\author{Amir Hajiyavand$^1$,}
\author{Hyunyoung Kim$^3$,}
\author{Claire Dancer$^4$,}
\author{Karl Dearn$^{1,*}$,}
\author{Mingchao Liu$^{1,*}$}


\begin{affiliations}

1. Department of Mechanical Engineering, University of Birmingham, Birmingham, B15 2TT, UK\\
2. School of Engineering, Newcastle University, Newcastle upon Tyne, NE1 7RU, UK\\
3. School of Computer Science, University of Birmingham, Birmingham B15 2TT, UK\\
4. School of Metallurgy and Materials, University of Birmingham, Birmingham B15 2TT, UK

$*$ Corresponding authors: weicheng.huang@newcastle.ac.uk (W.H.); k.d.dearn@bham.ac.uk (K.D.); \\ m.liu.2@bham.ac.uk (M.L.). \\

\end{affiliations}



\begin{abstract}

Programmable folding of elastic sheets typically relies on predefined flexible creases or active materials-enabled hinges, which lack intrinsic bistability and limit reprogrammability within a single structure. 
Here, we present a dimple-encoded origami platform that converts bistable dimple snapping into spatially addressable hinges with prescribed folding angles in a continuous sheet. This interaction-enabled mechanism enables the design of distributed hinge networks through the arrangement and selective inversion of dimples.
We establish folding-angle design charts that can be directly used to select local dimple arrangements for target fold angle, forming a practical hinge library without altering the underlying unit geometry. Using this approach, a single dimpled sheet can be reprogrammed to realize multiple distinct configurations, such as triangle, square, and pentagon shapes.
We further extend the method to flat-to-3D morphing of polyhedral origami and validate the results through experiments and finite element simulations. As demonstrations, we realize self-supporting cubic shells with enhanced impact resistance and partially deployable cube configurations that remain stable upon opening, highlighting their potential for protective enclosures and deployable architectural structures. The proposed strategy provides a fabrication-friendly route to reprogrammable shape-morphing and adaptive mechanical systems.

\end{abstract}


\section{Introduction}

Transforming planar sheets into three-dimensional (3D) structures through origami-inspired design has become a promising paradigm for creating deployable devices, reconfigurable metamaterials, and shape-morphing systems across a wide range of length scales \cite{meloni2021engineering,zhu2024large,misseroni2024origami}. By encoding geometry into a planar sheet, complex 3D morphologies and programmable mechanical functions can be generated through structural transformation, reducing the need for motors, bulky mechanisms, and heavy rigid assemblies. Such an approach offers lightweight solutions for robotics, wearable systems, biomedical devices, aerospace deployables, and protective packaging \cite{rus2018design,bolanos2018origami,yang2023morphing,yao2024origami}.

A central requirement in these applications is the ability to control folding behaviour through hinge design, namely to prescribe both the location and the magnitude of folding in a predictable and robust manner \cite{delimont2015evaluating,wagner2020hinges,miyazawa2023design,huang2024integration}. In most existing approaches, hinges are realized through predefined creases, compliant regions, or multi-material architectures that impose fixed folding pathways \cite{dureisseix2012overview,leanza2024active}. As a result, the folding behaviour is largely determined at the fabrication stage, and a given sheet can typically realize only a single target folded configuration.

While multistability has been explored as a route to enable reconfiguration, existing origami systems predominantly achieve multistability through global geometric constraints, leading to multiple stable configurations at the structure level \cite{waitukaitis2015origami,melancon2021multistable,liu2022triclinic,zhang2025reprogrammable}. In contrast, the design of hinges that are themselves multistable and capable of producing prescribed and tunable folding angles remains largely unexplored \cite{iniguez2022rigid}. The absence of such hinge-level multistability limits the ability to achieve reprogrammable folding within a single structure, where multiple distinct configurations can be accessed in a controlled manner.

Bistable elements offer a promising route to overcome these limitations, as they possess multiple stable equilibria separated by an energy barrier, enabling reconfiguration through snap-through and stable shape retention without continuous input \cite{cao2021bistable,chi2022bistable}.
Bistability arises in a wide variety of elastic structures, ranging from beams \cite{qiu2004curved} and ribbons \cite{huang2024integration} to shells \cite{taffetani2018static}. Among these, thin shells—such as spherical dimples—serve as canonical bistable units that can invert and remain stable due to geometric constraints. Arrays of such elements have therefore been widely explored for morphable sheets and mechanical metamaterials \cite{seffen2006mechanical,faber2020dome,liu2023snap,albertini2024mechanical}.
However, translating these bistable units into origami-style hinge systems with prescribed and predictable folding angles remains challenging. Existing dimpled sheets typically exhibit distributed curvature rather than localized hinge-like folding, while interactions between neighboring units introduce mode competition and sensitivity to spacing and arrangement, thereby complicating predictive design.

\begin{figure}[htbp]
    \centering
    \includegraphics[width=1.0\textwidth]{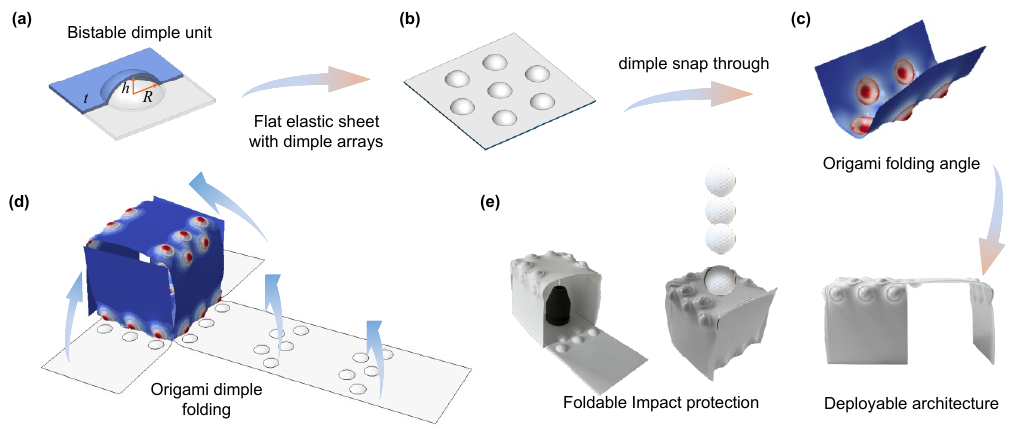} 
    \caption{\textbf{Dimple-encoded reprogrammable origami: concept and design principle.}
    (a) A bistable dimple unit defined by geometric parameters ($t$, $h$, $R$).
    (b) Arrays of dimples embedded in a flat sheet.
    (c) Interactions between neighboring snapped dimples generate effective hinge lines, defining a characteristic folding angle.
    (d) These interaction-induced hinges can be spatially organized through the placement and selective inversion of dimples to guide origami-like folding of the sheet.
    (e) The resulting dimple-encoded origami enables reprogrammable shape-morphing, allowing different target configurations to be achieved from the same sheet, with potential applications in deployable and adaptive structures.}
    \label{fig:dimple_origami}
\end{figure}

Here, we address this challenge by introducing a dimple-encoded multistable origami platform that leverages interactions between bistable dimples to realize spatially addressable hinges with programmable folding angles (Figure~\ref{fig:dimple_origami}).
By organizing dimples within a continuous sheet and selectively activating them through inversion, local bistable units are transformed into controllable hinge-like responses that guide folding. This strategy enables reprogrammable shape-morphing within a single structure, allowing multiple distinct configurations to be achieved from the same sheet.

Overall, this work establishes a design framework that leverages interactions between local bistable elements to generate programmable hinge networks with predictable folding, enabling scalable and fabrication-friendly reprogrammable morphing systems and deployable architectures.


\section{Results and Discussion}

\subsection{Dimple-Encoded Origami: Concept and Interaction-Enabled Hinges}
\label{sec:2.1}

We begin by introducing the concept of dimple-encoded origami, in which bistable dimples embedded in a continuous sheet are used to define spatially programmable hinge regions (Figure~\ref{fig:dimple_origami}). Each dimple acts as a local bistable unit that can be selectively inverted, generating localized deformation in the surrounding sheet (Figure~\ref{fig:dimple_origami}a). When multiple dimples are arranged in proximity, their interactions give rise to effective hinge-like behaviour, enabling controlled folding along prescribed directions (Figure~\ref{fig:dimple_origami}b,c).

This interaction-enabled mechanism provides a route to encode folding behaviour directly into the sheet through the spatial arrangement and activation of dimples (Figure~\ref{fig:dimple_origami}d). In contrast to conventional crease-based approaches, where hinge properties are fixed once fabricated, the present strategy allows hinge regions to be reconfigured by selectively inverting different subsets of dimples. As a result, a single sheet can access multiple distinct configurations, establishing the basis for reprogrammable shape-morphing.

Collectively, these interaction-enabled mechanisms allow bistable dimples to be organized into spatially distributed arrays whose selective activation defines programmable hinge networks within a continuous sheet. This enables controlled global shape transformations as well as reconfiguration between multiple stable states using the same structure.
Beyond geometric morphing, such interaction-enabled hinge networks provide a foundation for functional systems, including deployable enclosures and spatial structures that remain stable without continuous external support (Figure~\ref{fig:dimple_origami}e). These capabilities are explored in the following sections through quantitative design rules and representative demonstrations.

\subsection{Interaction-Guided Folding and Angle Design}
\label{sec:2.2}

\begin{figure}[!t]
    \centering
    \includegraphics[width=1.0\textwidth]{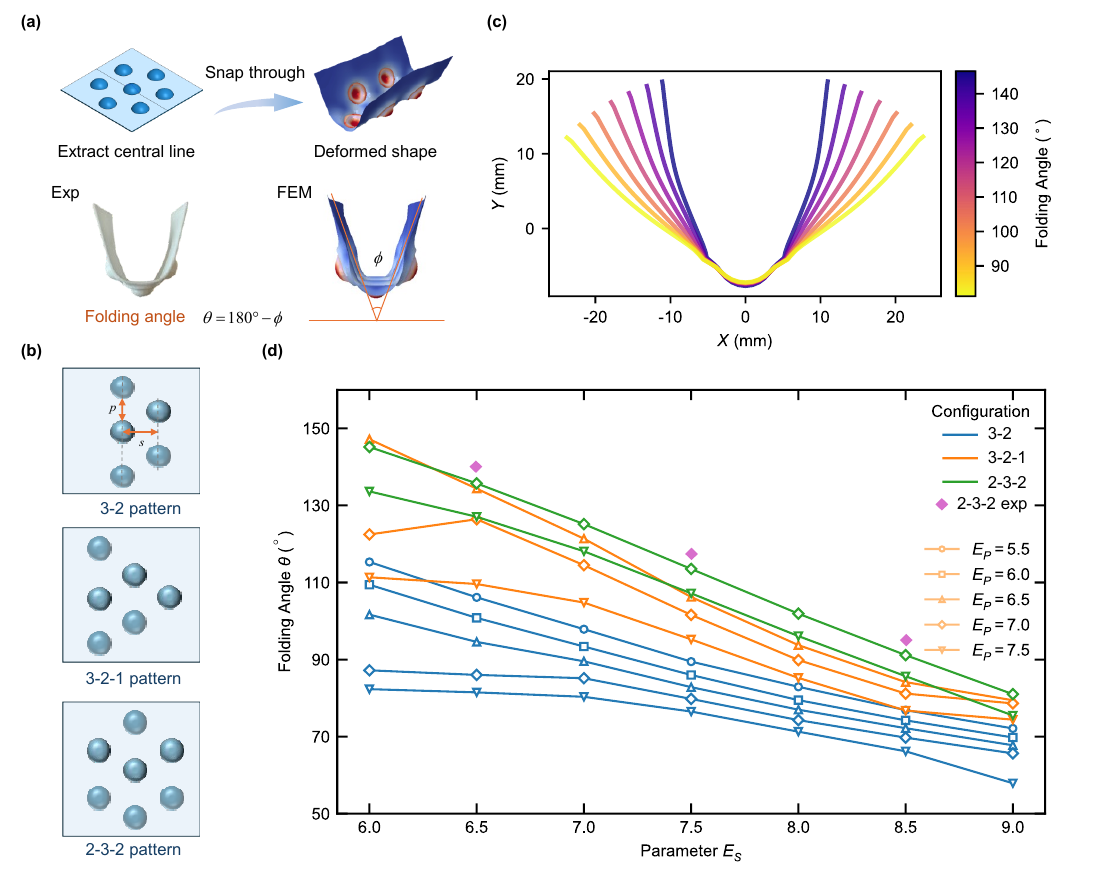}
    \caption{\textbf{Dimple interaction-enabled hinges and folding-angle design charts.}
    (a) Extraction of the hinge opening angle $\phi$ from the centerline of the deformed dimpled sheet, and the definition of the folding angle $\theta = 180^\circ - \phi$.
    (b) Definitions of the longitudinal pitch $p$ and transverse spacing $s$ (nondimensionalized as $E_p = p/l_p$ and $E_s = s/l_p$), and representative array topologies (`3--2', `3--2--1', and `2--3--2').
    (c) Representative deformed centerlines corresponding to different folding angles for a fixed array configuration (2--3--2 pattern) and representative parameters.
    (d) Design charts showing the folding angle $\theta$ as a function of $E_s$ for different values of $E_p$ and array configurations.}
    \label{fig:hinge_library}
\end{figure}

As illustrated in Figure~\ref{fig:hinge_library}a, hinge-like folding can be induced by arranging multiple bistable dimples within a continuous sheet and selectively activating them through inversion. Interactions between neighboring snapped dimples give rise to localized hinge regions, characterized by a well-defined folding angle $\theta$.

The key to programmable hinge behaviour therefore lies in controlling these dimple–dimple interactions. When a dimple snaps, the induced deformation localizes within a narrow region where bending and stretching compete \cite{liu2023snap}, introducing a natural interaction length scale that governs how nearby dimples couple.

A simple scaling argument captures this characteristic length. The competition between bending, which favors smooth curvature, and stretching, which penalizes in-plane deformation, leads to a localized boundary layer with a characteristic size \cite{liu2023snap}
\begin{equation}
\ell_b \sim \sqrt{tR},
\end{equation}
where $t$ is the sheet thickness and $R$ is the dimple radius. This characteristic length, also referred to as the Pogorelov length scale \cite{pogorelov1988bendings,gomez2016shallow}, provides a physical measure of the interaction range: strong coupling occurs when spacings are comparable to $\ell_b$, whereas weak coupling arises when dimples are well separated.

Based on this interaction length, we parameterize dimple arrays using the longitudinal pitch $p$ and transverse spacing $s$ (Figure~\ref{fig:hinge_library}b), normalized as
\begin{equation}
E_p = \frac{p}{\ell_b}, \qquad E_s = \frac{s}{\ell_b}.
\end{equation}
This nondimensional description collapses geometric effects into a compact interaction-based design space, enabling different array topologies to be compared on equal footing.

Since a single row of dimples fails to produce well-defined and controllable folding (Figure~S1), we consider three representative multi-row patterns, namely `3--2', `3--2--1', and `2--3--2', as shown in Figure~\ref{fig:hinge_library}b. The corresponding design spaces for each pattern, in which all snapped dimples remain bistable and the resulting deformation localizes into a well-defined hinge, are mapped in Figure~S2.To experimentally validate the folding angle, we employed an elastic sheet measuring 60~mm $\times$ 60~mm. The geometric parameters of the dimple were set to a radius $R$ of 5~mm and a height $h$ of 3.5~mm. The design rationale for the chosen dimple geometry is presented in Figure~S2a.

Using the folding angle $\theta$ extracted from the deformed configuration obtained via finite element method (FEM) simulations (Figure~\ref{fig:hinge_library}c) and experiments (Figure~S3), we construct angle design charts that map $(E_p, E_s)$ to the resulting hinge response for each topology (Figure~\ref{fig:hinge_library}d). Across all configurations, two consistent trends emerge: increasing either $E_s$ or $E_p$ weakens dimple–dimple coupling and leads to shallower folds (smaller $\theta$), reflecting reduced interaction-induced incompatibility.

The two spacing directions and the hinge topology play distinct yet complementary roles in design. The transverse spacing $E_s$ governs cross-column coupling and can induce qualitative changes in deformation mode, often accompanied by distributed bending or warping (Figure~S2b-d), whereas the longitudinal pitch $E_p$ provides a continuous and predictable parameter for tuning the folding angle within a given mode.
The hinge topology further sets the accessible angle range: the `3--2' configuration yields smaller folding angles, whereas `3--2--1' and `2--3--2' enable larger $\theta$ and a wider range, particularly at smaller $E_s$. At larger $E_s$, the responses of different topologies converge, consistent with weakened coupling in the weak-interaction regime.
These results establish a practical design strategy: $E_s$ and topology are first selected to ensure a stable deformation mode, after which $E_p$ is used to continuously tune the target folding angle.

Taken together, these results establish a geometry-driven design framework for programmable hinge behavior. The resulting angle design charts (Figure~\ref{fig:hinge_library}d) thus function as a practical hinge library, enabling prescribed folding angles to be achieved without modifying the underlying bistable units.

\subsection{Reprogrammable Multi-Hinge Origami Structures}
\label{sec:2.3}

With the folding angle design charts established, we next demonstrate how these calibrated hinges can be composed to program system-level origami morphing. The key idea is to treat a single strip as a sequence of quasi-rigid segments connected by localized dimple-encoded hinges, where each hinge is assigned a prescribed folding angle by selecting its local array topology and interaction parameters $(E_p, E_s)$ from the hinge library given in Section \ref{sec:2.2}.

\begin{figure}[!h]
    \centering
    \includegraphics[width=1.0\textwidth]{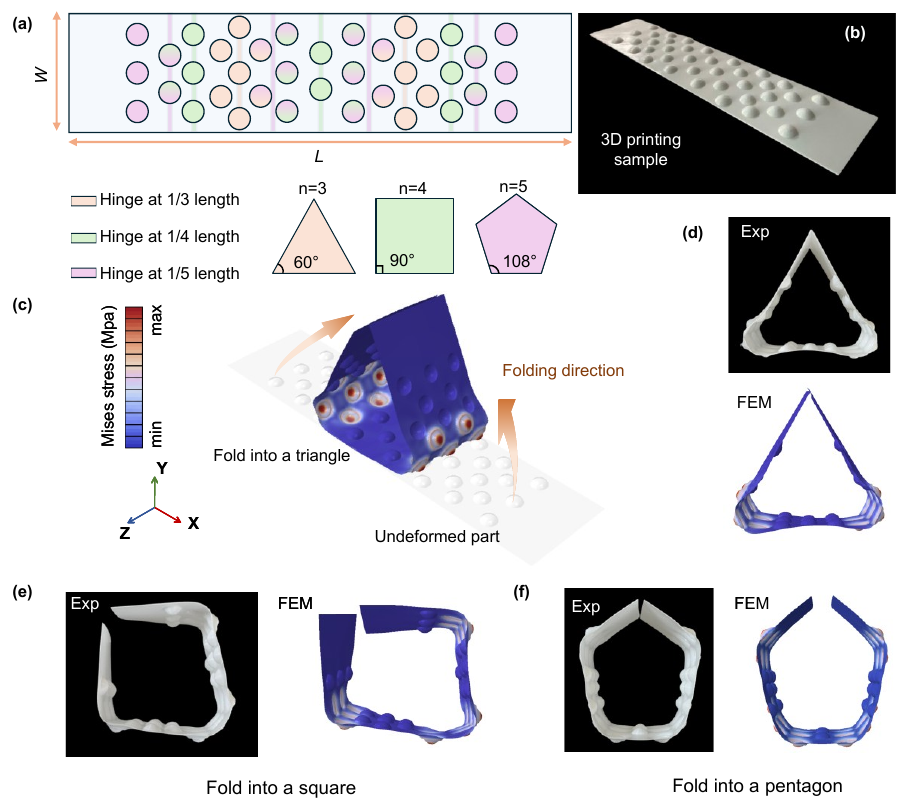} 
    \caption{\textbf{Dimple-encoded multi-hinge strip for reprogrammable shape-morphing.}
    (a) (Top) Schematic of a single elastic strip patterned with bistable dimple arrays, where different segments encode hinge regions with prescribed folding angles. (Bottom) Corresponding hinge-placement rule for regular polygonal shapes, where the required turning angle is the exterior angle $\psi_n = 360^\circ/n$ (triangle: $120^\circ$, square: $90^\circ$, pentagon: $72^\circ$; the corresponding interior angles are $60^\circ$, $90^\circ$, and $108^\circ$).
    (b) 3D-printed prototype of the dimple-encoded strip.
    (c) Demonstration of the folding process into an equilateral triangle through FEM simulation.
    (d) Folding into an equilateral triangle, with comparison between experimental realization (Exp) and numerical prediction (FEM).
    (e) Reconfiguration of the same strip into a square.
    (f) Reconfiguration into a regular pentagon, further demonstrating reprogrammable shape-morphing within a single structure.}
    \label{fig:Multihinge_strip}
\end{figure}

Figure~\ref{fig:Multihinge_strip} illustrates the design and implementation of a single strip that can be reprogrammed into multiple target shapes. In the strip layout, colored bands indicate the hinge locations, each encoding a distinct target turning angle (Figure~\ref{fig:Multihinge_strip}a). For a regular $n$-gon frame, the turning angle at each vertex is given by the exterior angle \cite{kontorovich2016turn}
\begin{equation}
\psi_n = \frac{360^\circ}{n},
\end{equation}
which can be realized by assigning hinge segments with $\theta \approx \psi_n$ using the calibrated $(E_p, E_s)$–$\theta$ relation. Notably, this programming is achieved without altering the bistable unit geometry: only the local spacing and topology are varied, making the approach fabrication-friendly and scalable.The elastic strip used in both experiments and simulations has dimensions of length $L = 250$~mm and width $W = 60$~mm.

Using this workflow, multiple hinge sets are encoded along the same strip to target different polygonal configurations. A representative strip is fabricated by 3D printing, as shown in Figure~\ref{fig:Multihinge_strip}b. Figure~\ref{fig:Multihinge_strip}c demonstrates that, by snapping selected dimples, the hinges at one-third of the strip length are activated, folding the strip into a triangular configuration.

Furthermore, as shown in Figure~\ref{fig:Multihinge_strip}d--f and Movie~S1, the same strip can be reconfigured into a triangle, square, and pentagon by selectively inverting different subsets of dimples. Experiments and FEM simulations show good qualitative agreement, confirming that the interaction-guided hinge library enables predictable and repeatable reprogramming within a single structure. Each configuration is maintained as a stable state without external constraint due to the intrinsic bistability of the hinges.

Two considerations are important for reliable composition. First, hinge segments must operate within the bistable regime to avoid relaxation and unintended spring-back (Figure~S2). Second, sufficient spacing between neighboring hinges is required to minimize interaction between adjacent boundary layers, which may otherwise distort the prescribed angles and accumulate geometric error.

Together, these results demonstrate that the interaction-guided hinge library is not only descriptive but also constructive: it enables distributed angle encoding along a single structure to achieve multiple distinct configurations on demand. In the next section, we extend this concept from planar frames to fully three-dimensional polyhedral origami shells.

\subsection{Flat-to-3D Polyhedral Origami}
\label{sec:2.4}

Having established reprogrammable folding within planar strips, we next demonstrate that the same interaction-guided hinge library can be extended to construct fully three-dimensional origami structures from a flat sheet. The central idea is to encode spatially distributed hinges directly into a planar net such that the target polyhedral geometry emerges upon activation, without the need for discrete joints, multi-material interfaces, or external assembly.

Our approach is constructive. A target polyhedron is first unfolded into a planar net, and each fold line is assigned a localized dimple-encoded hinge with a prescribed folding angle selected from the design charts. The faces remain comparatively rigid, while deformation is confined to the programmed hinge regions, allowing the global three-dimensional shape to be dictated entirely by local geometry. In this way, complex shell structures can be realized through a single-material, geometry-driven design framework.

\begin{figure}[!t]
    \centering
    \includegraphics[width=1.0\textwidth]{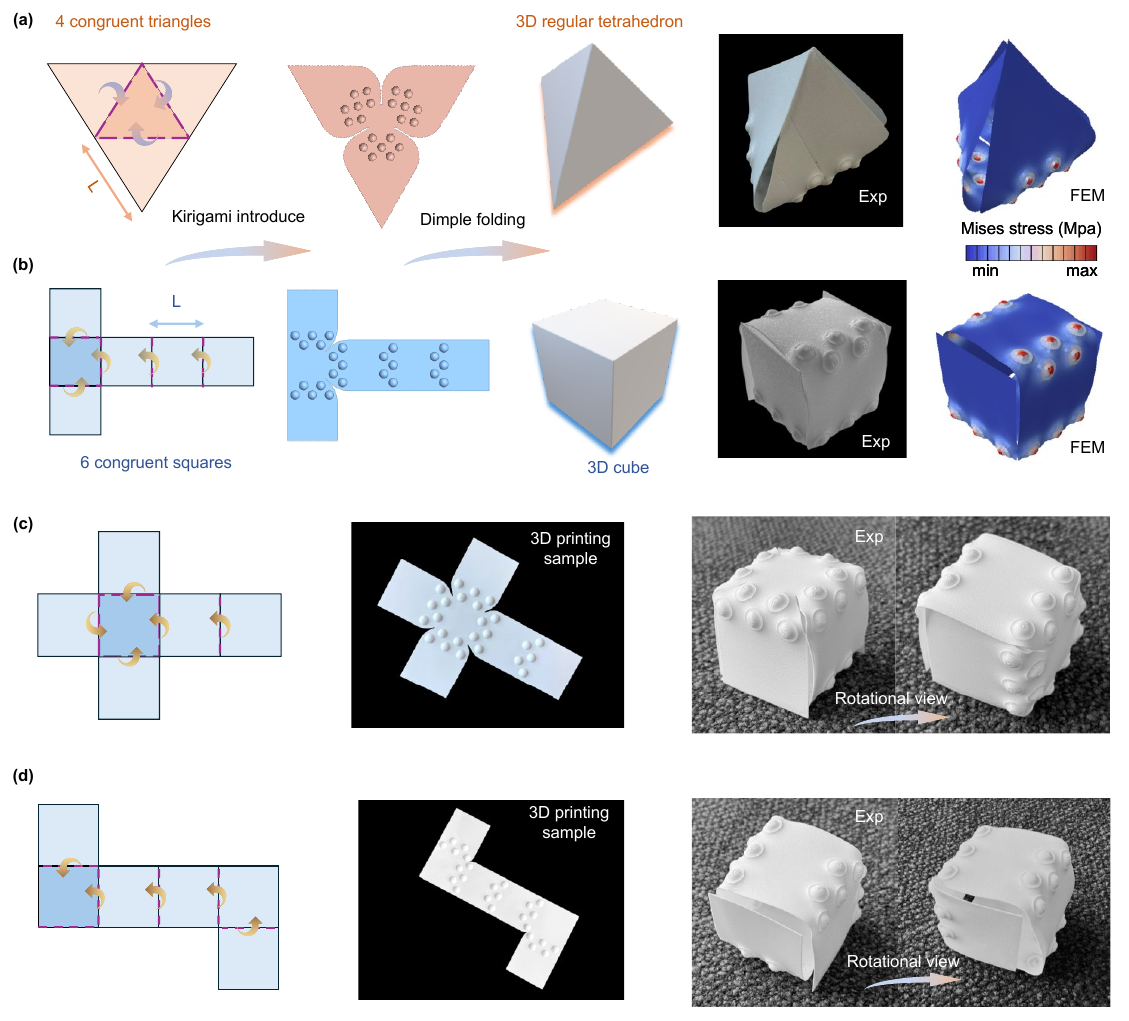}
    \caption{\textbf{Flat-to-3D polyhedral shells programmed by dimple-encoded hinges.}
    (a) Planar net of four congruent triangles, where dimple hinges with angles selected from the calibrated hinge library (Figure~2) are encoded and kirigami cuts are introduced to enable folding into a regular tetrahedron; experimental realization (Exp) and numerical prediction (FEM) are shown for comparison and demonstrate agreement in the assembled shell and deformation localization near hinge regions.
    (b) Planar net of six congruent squares folds into a cube, with validation from both Exp and FEM results.
    (c, d) Alternative cube net layouts fabricated by 3D printing and their corresponding folded cube enclosures in experiments, shown in different rotational views, illustrating that different planar layouts can realize the same target polyhedron under the same hinge-selection workflow.}
    \label{fig:polyhedra}
\end{figure}

Figure~\ref{fig:polyhedra} illustrates this concept through representative demonstrations. A net composed of four congruent triangular panels folds into a regular tetrahedron (Figure~\ref{fig:polyhedra}a), providing a minimal example that validates the generality of the approach across different folding angles. Kirigami cuts are introduced at the hinge edges to suppress out-of-plane warping \cite{liu2020tapered} (Figure~S4). In the flat-to-3D folding experiments and simulations, the planar net of the regular tetrahedron consists of four congruent equilateral triangles with a side length of $L = 120$~mm, resulting in a total footprint length of 240~mm. For the hexahedron (3D cube), the planar layout comprises six congruent squares, each characterized by a side length of $L = 60$~mm.
Next, a net of six square panels is programmed to assemble into a cubic shell, forming a robust, self-supporting enclosure (Figure~\ref{fig:polyhedra}b). 3D-printed experiments and FEM simulations show consistent folding pathways and final geometries, with deformation strongly localized at the hinge regions, thereby confirming the predictive capability of the hinge library. The folding of the printed tetrahedron and cube is shown in Movie~S2.

Beyond geometric assembly, this platform enables functional three-dimensional systems. The resulting polyhedral shells are intrinsically stable due to the bistability of the hinges, allowing them to maintain their shape without continuous external input. Moreover, different planar nets leading to the same target polyhedron can be encoded within the same design framework (Figure~\ref{fig:polyhedra}c,d), offering flexibility in deployment pathways and accessible intermediate configurations.
This ability to program both the final shape and the transformation pathway highlights the potential of dimple-encoded origami for deployable enclosures, protective structures, and scalable architectural systems. From a design perspective, the planar net introduces an additional programming dimension beyond hinge-angle selection: while the hinge library prescribes local folding angles, the net geometry governs the global folding pathway and determines the set of accessible configurations and functional states.

These results demonstrate that the interaction-guided hinge library extends naturally from planar morphing to spatial assembly, providing a unified route to fabricate complex, self-supporting three-dimensional origami structures from a single sheet.

\subsection{Functional Demonstrations: Impact Protection and Deployable Spatial Structures}
\label{sec:2.5}

\begin{figure}[!h]
    \centering
    \includegraphics[width=1.0\textwidth]{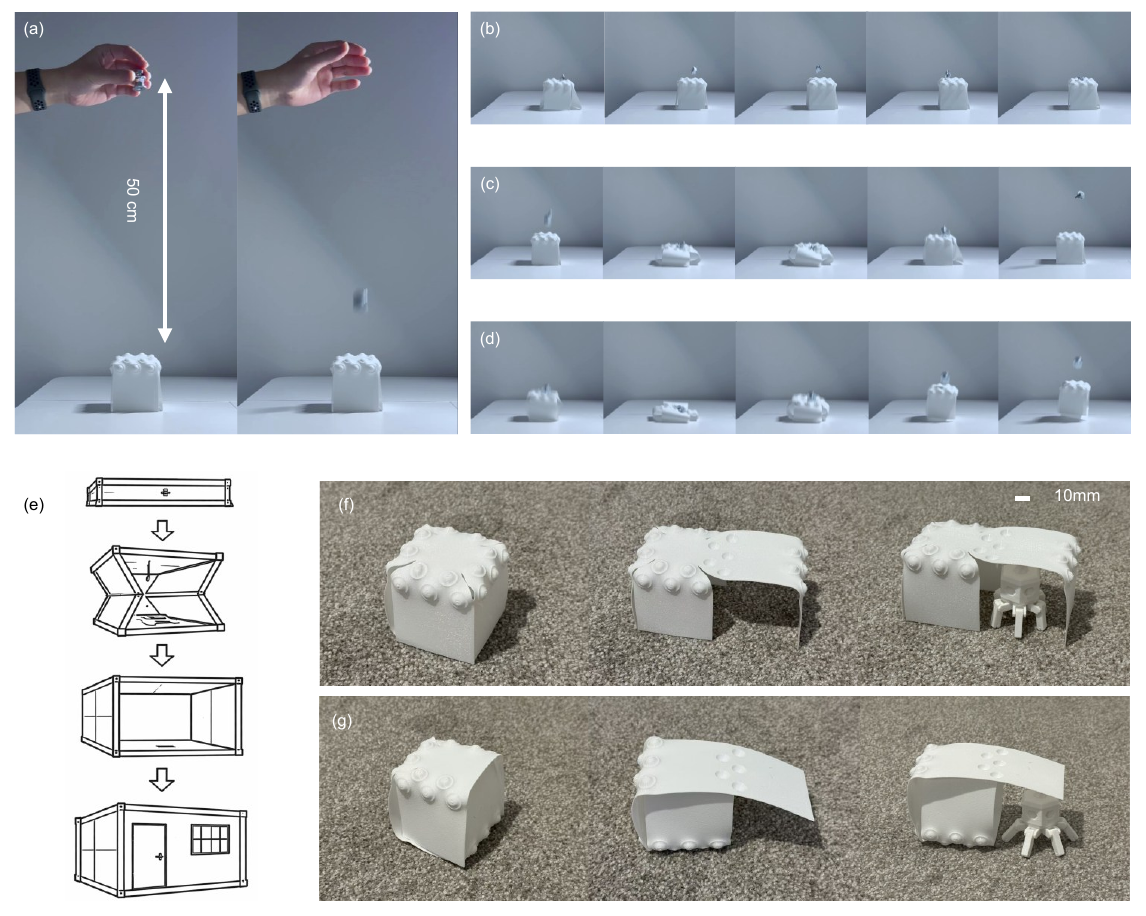}
    \caption{\textbf{Functional demonstrations of dimple-encoded deployable origami cubes: impact protection and scalable spatial configurations.}
    (a) Schematic of the drop-impact test.
    (b–d) Impact-protection demonstration: sequential frames from a single drop-impact experiment showing the response of the pre-folded origami cube under increasing impact energies. The structure deforms upon impact while maintaining an enclosed volume, thereby protecting the payload.
    (e) Schematic illustration of a deployable architectural concept, highlighting multiple stable states from fully enclosed to partially deployed configurations.
    (f) Partially deployed configuration forming a stable and accessible spatial structure.
    (g) Alternative partially deployed configuration illustrating net-dependent geometries for deployable architectural applications.}
    \label{fig:functional_cube}
\end{figure}

The dimple-encoded origami serves not only as a geometric demonstration but also as a functional structure with intrinsic multistability and reprogrammability, whose behaviour can be tailored through the choice of planar net and folding pathway. Figure~\ref{fig:functional_cube} summarizes two complementary functionalities enabled by the same dimple-programmed origami cube.

First, the origami cube provides impact protection for delicate payloads \cite{sareh2018rotorigami}. A payload placed inside the programmed cube remains shielded under controlled drop/impact tests (Figure~\ref{fig:functional_cube}a–d and Movie~S3; details in the Experimental Section), demonstrating that the bistable, hinge-programmed origami shell can maintain a protective cavity and mitigate impact transmission. This protection arises from two coupled mechanisms: the closed polyhedral geometry distributes loads across multiple faces and hinges, avoiding localized failure, while the bistable dimple hinges provide local buffering through deformation, reducing direct load transmission to the payload. Notably, as the impact load increases from Figure~\ref{fig:functional_cube}b to d, the resulting deformation becomes progressively larger; however, in all cases, the structure fully recovers its original shape without external intervention, demonstrating its impact resilience.

Second, the dimple-encoded origami structure provides a model system for deployable architectural designs \cite{wang2025origami} (Figure~\ref{fig:functional_cube}e). As illustrated in Figure~\ref{fig:functional_cube}f,g, the programmed cube can transition between fully enclosed and partially deployed configurations through the selective inversion of bistable dimple hinges, with each state remaining intrinsically stable without external support. This multistability enables controllable spatial reconfiguration within a single structure, allowing the degree of enclosure to be actively programmed. Although demonstrated at a small scale, the underlying design principles are inherently scalable, suggesting a pathway toward geometry-driven, reconfigurable architectural structures while requiring further advances in manufacturing and materials for large-scale implementation.

Together, these results demonstrate that interaction-guided bistable dimples constitute a constructive design language: calibrated local hinge behaviour can be composed to program global flat-to-3D assembly, while net geometry enables functional adaptability within a single deployable system.

\section{Conclusion}

In summary, we developed a dimple-encoded bistable origami platform that converts snap-through of a single bistable unit into spatially addressable hinges with programmable folding angles in a continuous sheet. By exploiting the interaction length scale associated with shallow-shell boundary-layer localization, we introduced a compact nondimensional design space for dimple coupling and established a robust fold-angle metric, enabling quantitative angle design charts that can be directly inverted to select local dimple arrangements for prescribed folding responses.
This interaction-guided framework elevates bistable dimples from a qualitative morphing mechanism to a predictive and constructive design tool. Using this approach, we demonstrated reprogrammable multi-hinge strips capable of accessing multiple polygonal configurations within a single structure, and extended the same strategy to flat-to-3D assembly of polyhedral shells, including a deployable cube and a tetrahedron. Beyond geometric reconfiguration, the cube exhibits functional capabilities, including impact protection and multistable, partially deployed spatial configurations.
Overall, this work establishes a fabrication-friendly route to programmable and reprogrammable shape-morphing that bridges local bistable interactions and system-level structural functionality. The proposed design paradigm opens opportunities for reconfigurable mechanical metamaterials, protective enclosures, and scalable deployable structures inspired by grippers and architectural systems.

\section{Experimental Section}
\threesubsection{Sample Fabrication}\\
All experimental specimens were fabricated using Bambu TPU 90A filament (Bambu Lab) with a Raise3D E2 fused deposition modeling (FDM) printer. The printed TPU has an elastic modulus of $E = 8$~MPa and a Poisson’s ratio of $\nu = 0.45$. Prior to printing, the filament was dried at 70~$^\circ$C for 7~h. A 0.4~mm nozzle and a layer height of 0.1~mm were used. Printing was performed at a speed of 20~mm~s$^{-1}$, with the build-plate temperature set to 45~$^\circ$C and the nozzle temperature to 223~$^\circ$C.

\threesubsection{Impact Protection and Deployable architectural demonstration}\\
\textit{Drop-impact test.}
Impact protection was evaluated using a vertical drop test, as illustrated in Figure~\ref{fig:polyhedra}a. Pre-folded origami cubes were placed on a rigid substrate, and a rigid impactor was released from controlled heights to generate different impact energies. To evaluate the impact resistance, impact tests were conducted by releasing three distinct masses $m = 9.28$, $21.06$, and $30.28$~g from a constant drop height of $h = 50$~cm. According to the gravitational potential energy formula $E = mgh$, these experimental conditions correspond to impact energies of $0.046$, $0.103$, and $0.149$~J, respectively.
The impact process was recorded using a high-speed camera, from which sequential frames were extracted for analysis (Figure~\ref{fig:functional_cube}b–d). The deformation of the structure and the response of the enclosed payload were qualitatively assessed. Each test was repeated at least three times to ensure reproducibility.

\textit{Deployable configurations.}
The reprogrammable folding capability was further demonstrated through partially deployed configurations of the origami cube (Figure~\ref{fig:functional_cube}e–g). Starting from the fully enclosed state, selected hinges were actuated to access intermediate stable configurations. The resulting structures remain self-supporting and exhibit different accessible geometries depending on the underlying net design.

\threesubsection{Modelling and Simulation}\\
3D CAD models were created in SolidWorks 2024. FEM analyses were performed in Abaqus/Standard on the University of Birmingham BlueBEAR high-performance computing (HPC) cluster. All cases were modeled using S4R shell elements and a consistent implicit dynamic (quasi-static) procedure to capture snap-through and post-buckling responses. A three-step loading and relaxation protocol was adopted. In Step~1, all boundary edges of the sheet were fully constrained and a pressure load was applied to the dimple surfaces with a linear ramp to trigger inversion. In Step~2, the pressure load was removed while maintaining the edge constraints to allow the structure to reach an equilibrium configuration. In Step~3, all boundary constraints were released to obtain the final free (relaxed) deformed state.The material behaviour was described using a compressible Neo-Hookean hyperelastic model with parameters $C_{10}=1.4286$ and $D_{1}=0.15$. For all simulations, energy-based convergence was checked to ensure stable solutions. Mesh convergence was verified by mesh-independence studies prior to the parameter sweeps. 
\medskip

\textbf{Supporting Information} \par 
Supplementary information for this article is provided in the attached file.

\medskip
\textbf{Acknowledgements} \par 
Q.Z. acknowledges the scholarship support from the School of Engineering at the University of Birmingham. M.L. acknowledges the start-up funding from the University of Birmingham, UK.

\textbf{Conflicts of Interest}\par
The authors declare no conflict of interest.

\textbf{Data Availability Statement}\par              
The data that support the findings of this study are available from the corresponding author upon reasonable request.
\medskip

%


\bibliographystyle{MSP}
\bibliography{references}



\end{document}